\documentclass[twocolumn,superscriptaddress,nofootinbib,10pt]{revtex4-1}

\usepackage{graphicx,xcolor,amsmath}
\usepackage{hyperref}
\usepackage{bm}
\usepackage{cancel} % to write SUSY-breaking

\definecolor{c1}{rgb}{0, 0, 1}
\definecolor{c2}{rgb}{0.8, 0,0}
\definecolor{c3}{rgb}{0, 0.5, 0}
\definecolor{c4}{rgb}{0.5, 0, 0.5}
\definecolor{c5}{rgb}{0.9, 0.4, 0}
\definecolor{c6}{rgb}{1., 0.1, 0.8}
\definecolor{c7}{rgb}{0.58, 0.58, 0.58}
\definecolor{c8}{rgb}{0.53, 0.28, 0.}
\definecolor{c9}{rgb}{0, 0, 0}

\definecolor{myprimary}{HTML}{D9661F} % gold
\definecolor{mysecondary}{HTML}{003262} % blue
\definecolor{mytertiary}{HTML}{00A598}%FDB515%EE1F60%3B7EA1%ED4E33%6C3302

\usepackage{tikz}
\usepackage{tkz-euclide}
\usetkzobj{all}
\usetikzlibrary{backgrounds}
\usetikzlibrary{decorations.pathmorphing}	% To create gluons and vector bosons
\tikzset{
    v/.style={decorate, decoration={snake, segment length=3mm, amplitude=0.75mm}, draw},
    f/.style={draw=black, postaction={decorate},
        decoration={markings,mark=at position .6 with {\arrow[very thick]{latex}}}},
    fb/.style={draw=black, postaction={decorate},
        decoration={markings,mark=at position .4 with {\arrowreversed[very thick]{latex}}}},
    fnar/.style={draw=black},
    g/.style={decorate, draw=black,
        decoration={coil,amplitude=3pt, segment length=3.5pt}},
    s/.style={dashed,draw=black, postaction={decorate},
        decoration={markings,mark=at position .55 with {\arrow[very thick]{latex}}}},
    sb/.style={dashed,draw=black, postaction={decorate},
        decoration={markings,mark=at position .55 with {\arrowreversed[draw=black,very thick]{latex}}}},
    snar/.style={dashed,draw=black,line width =1.25pt},
}

\usepackage{epstopdf}
\usepackage{setspace}
\usepackage{feynmp-auto}
\newcommand{\of}[1]{\left( #1 \right)}
\newcommand{\sqof}[1]{\left[ #1 \right]}
\newcommand{\abs}[1]{\left| #1 \right| }

\begin{document}

\title{Dark Matter in Very Supersymmetric Dark Sectors}

\author{Avital Dery}
\email{avital.dery@weizmann.ac.il }\affiliation{Department of Particle Physics and Astrophysics,
Weizmann Institute of Science, Rehovot, Israel 7610001}
\author{Jeff A. Dror}
\email{jdror@lbl.gov}\affiliation{Department of Physics, University of California, Berkeley, CA 94720, USA}
\affiliation{Ernest Orlando Lawrence Berkeley National Laboratory, University of California, Berkeley, CA 94720, USA}
\author{Laurel Stephenson Haskins}
\email{laurel.haskins@mail.huji.ac.il }\affiliation{Racah Institute of Physics, Hebrew University of Jerusalem, Jerusalem 91904, Israel}
\author{Yonit Hochberg}
\email{yonit.hochberg@mail.huji.ac.il}\affiliation{Racah Institute of Physics, Hebrew University of Jerusalem, Jerusalem 91904, Israel}
\author{Eric Kuflik}
\email{eric.kuflik@mail.huji.ac.il}\affiliation{Racah Institute of Physics, Hebrew University of Jerusalem, Jerusalem 91904, Israel}

%\date{\today}

\begin{abstract}
If supersymmetry exists at any scale, regardless of whether it is restored around the weak scale, it may be a good symmetry of
the dark sector, enforcing a degeneracy between its lowest lying fermions and bosons. We explore the implications of this scenario for the early universe and dark matter, as well as the
corresponding signatures. In particular we show that the thermal history of the dark sector results
in co-decaying dark matter in much of the parameter space. This implies new phenomenological
signatures and presents a new way to discover high scale supersymmetry.

 \end{abstract}

\maketitle

%%%%%%%%%%%%%%%%%%%%%%%%%%%%%%%%%%%%%%%%%%%%%%%%%%

%%%%%%%%%%
\section{Introduction}\label{sec:intro}

The identity of dark matter (DM) has long been a mystery, as has the question of how its observed relic abundance came to be. The lack of observation of WIMP dark matter at operating direct detection experiments has led to a surge in activity in this field, with many proposed new theoretical ideas for dark matter beyond the WIMP (see, e.g., Refs.~\cite{Pospelov:2007mp,Hochberg:2014dra,Hochberg:2014kqa,Kuflik:2015isi,Carlson:1992fn,Pappadopulo:2016pkp,Farina:2016llk,Griest:1990kh,Dror:2016rxc,Dror:2017gjq}). Many new experimental ideas to detect such dark matter have similarly been proposed (see, e.g., Refs.~\cite{Essig:2015cda,Essig:2011nj,Essig:2012yx,Essig:2017kqs,Derenzo:2016fse,Hochberg:2016ntt,Hochberg:2015pha,Hochberg:2015fth,Hochberg:2017wce}).

Supersymmetry (SUSY) has 
%long 
been a focal point for particle physics research
since its conception,
though null results from the LHC have somewhat shifted the image of what a supersymmetric solution to the hierarchy problem would look like. For the past 35 years, the study of dark matter with supersymmetry has focused on the lightest superpartner (LSP) of the Minimal Supersymmetric Standard Model (MSSM) as a dark matter candidate, as it can be absolutely stable due to R-parity conservation~\cite{Goldberg:1983nd}.

In this work we take a different approach, and show that if there is a hidden sector that is approximately supersymmetric, with an interaction portal to the MSSM, then there are profound implications for the thermal history of dark matter.
Approximately supersymmetric sectors have been considered in the context of WIMP dark matter in Refs.~\cite{Cheung:2009qd,Baumgart:2009tn,Morrissey:2009ur,Cohen:2010kn}, and the collider physics implications of such sectors have been studied extensively in stealth supersymmetry~\cite{Fan:2011yu,Fan:2012jf,Fan:2015mxp}. 
Such sectors can undergo different thermal histories depending on the size of the coupling between the visible and dark sectors; as we will show, this results in co-decaying dark matter~\cite{Dror:2016rxc} in much of the parameter space. 

In the co-decaying dark matter framework, recently proposed in Ref.~\cite{Dror:2016rxc}, nearly degenerate particles of a dark sector are in chemical and thermal contact with each other, while the lighter of the states undergoes a slow, out-of-equilibrium, decay into Standard Model (SM) particles. Supersymmetric theories are therefore a natural habitat for the co-decaying mechanism, since the near-degeneracy of the dark particles is ensured by approximate supersymmetry within the hidden sector.

It is clear that supersymmetry, if it exists at some scale, is badly broken in the visible sector. It may, however, be nearly preserved in a hidden sector.
%
% Although supersymmetry, if it exists at some scale, is clearly badly broken in the visible sector, it may nonetheless be 
%it is nonetheless possible that the symmetry is 
%
%While it is clear that if Nature is supersymmetric at some scale then it is badly broken in the visible sector, it is nonetheless possible that the symmetry is nearly preserved in a hidden sector.
%
Supersymmetry breaking in a hidden sector can naturally be small if it is mediated from the supersymmetry breaking sector to the MSSM, but not directly to the dark sector, e.g., via gauge mediation. The dark sector then feels supersymmetry breaking through the portal with the MSSM and through supergravity effects. Provided that the portal is weak, and the gravitino mass is much smaller than the supersymmetric masses in the hidden sector, the dark sector states will remain approximately degenerate within their supersymmetric multiplets.

Conventional models of gauge-mediated supersymmetry breaking suffer from the gravitino problem, where the gravitino abundance would over-close the universe unless all the superpartners are close to the TeV scale or the scale of reheating is low~\cite{Moroi:1993mb,Hall:2013uga}. As we will show, in models undergoing co-decay, this tension is alleviated through the entropy dump of the hidden sector into the SM particles. This substantially reduces the predicted relic abundance of the gravitinos, thus opening up the possibility of gauge-mediated split supersymmetry~\cite{Cohen:2015lyp} with a higher reheat temperature.

This paper is organized as follows. In Section~\ref{sec:SUSY} we discuss some generalities of approximately supersymmetric sectors. Section~\ref{sec:co-decay} reviews the ingredients of co-decaying dark matter. An explicit model is presented in  Section~\ref{sec:model}, with Section~\ref{sec:constraints} studying the parameter space. We conclude with a discussion in Section~\ref{sec:discussion}.

%%%%%%%%%%%%
\section{Very supersymmetric sectors}
\label{sec:SUSY}
\begin{figure*}[th!]
\begin{center} 
\hspace*{1em}\raisebox{0.5em}{
\begin{tikzpicture}[line width=.45] 
    \draw[c2,dotted,very thick] (-2.5,3.5) -- (2.5,3.5) node[midway,below] {$ M $};
    \draw[c2,dotted,very thick] (-2.5,4) -- (2.5,4) node[midway,above] {$ M_{\rm pl} $};
    \draw[c2,dotted,very thick] (2.5,3.75) -- (0,0) node[midway,xshift=0.3cm,yshift=-0.2cm] {$ M_{\rm pl} $};
    \draw[c2,dotted,very thick] (-2.5,3.75) -- (0,0) node[midway,xshift=-0.2cm,yshift=-0.2cm] {$ \varepsilon  $};
    \draw[preaction={fill=white},fill=c1,opacity=0.4,very thick] (0,0) node[black,opacity=1] {\begin{tabular}{c}\large hidden \\ $\kappa , m $ \end{tabular}} circle (1.2cm); 
    \draw[preaction={fill=white},fill=c1,opacity=0.4,very thick] (-2.5,3.75) node[black,opacity=1] {\begin{tabular}{c}\large visible \\ $\alpha _i  , v$ \end{tabular}} circle (1.2cm);
    \draw[preaction={fill=white},fill=c1,opacity=0.4,very thick] (2.5,3.75) node[black,opacity=1] {\begin{tabular}{c}${\large \cancel{{\rm SUSY}}}$ \\ $  \sqrt{F}  $ \end{tabular}} circle (1.2cm);
    \vspace{1cm}
  \end{tikzpicture}
  }
~\hfill\includegraphics[width=.45\textwidth]{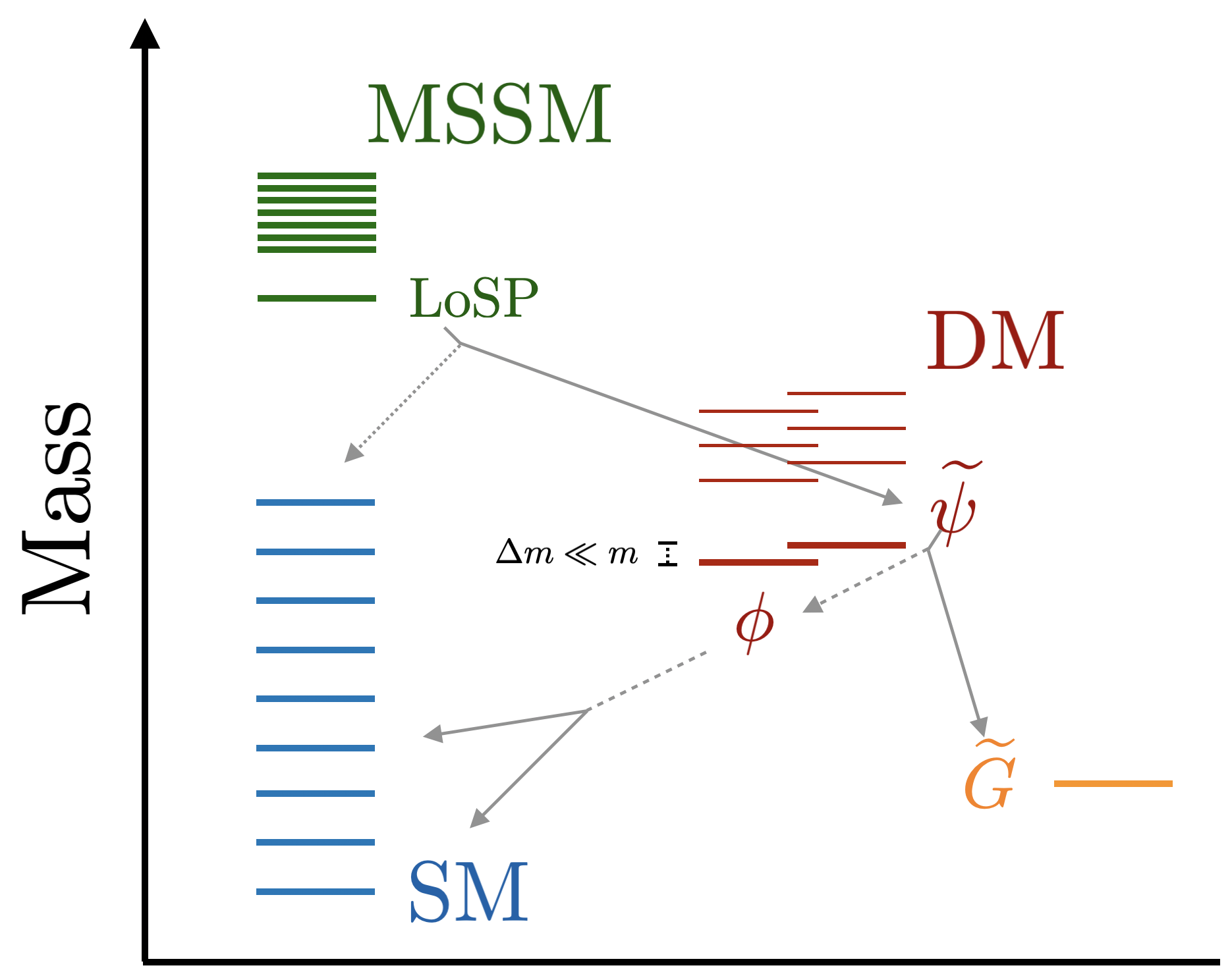}~\hspace{1em}
%~\hfill\includegraphics[width=.45\textwidth]{images/scheme.png}~\hspace{1em}
\caption{{\bf  Left:} Diagrammatic representation of the different sectors and their respective couplings. The hidden sector is connected to SUSY-breaking through the SM via $ M $ suppressed operators and directly through $ M_{\rm pl} $ suppressed operators.   {\bf Right:} Mass spectrum of particles we consider. 
}
\label{fig:sectors}
\end{center}
  \end{figure*}
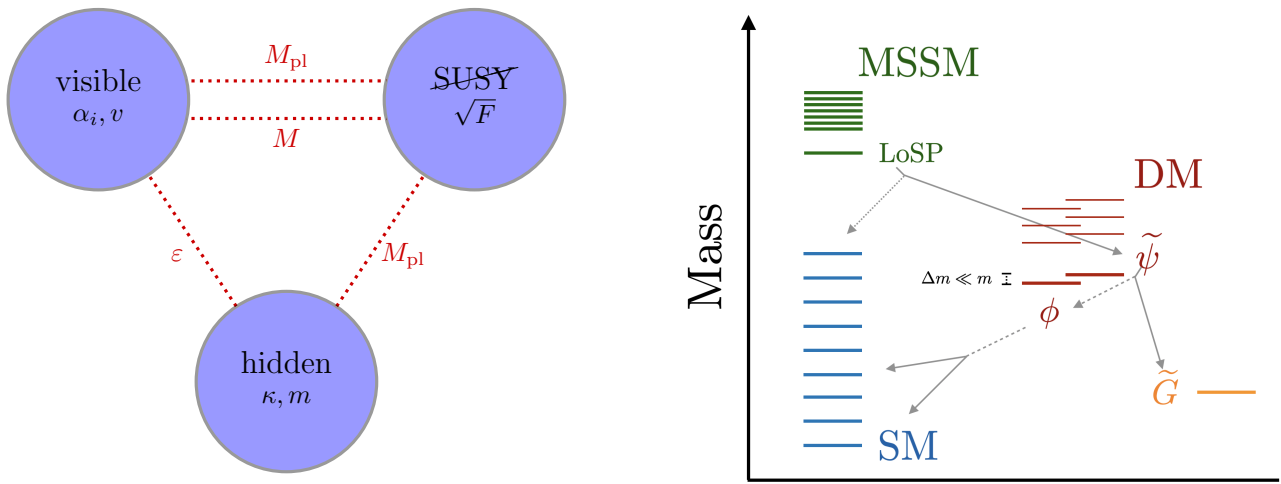

    We begin by summarizing the general features of having a {\em very supersymmetric} hidden sector, which we take to be a sector with small boson-fermion splittings within a supersymmetric multiplet. To best characterize the dynamics of such theories, we introduce the relevant scales and schematic shown in the left panel of Fig.~\ref{fig:sectors}. The hidden sector has a self-coupling $ \kappa$, mass $ m $, and is connected to the visible sector through a small coupling $ \varepsilon $.
The SUSY breaking is mediated to both the visible and hidden sectors through $ M_{\rm pl} $-suppressed operators and to the SM through its gauge interactions at a scale $ M $.

The thermal history of the dark sector and whether there exists a viable dark matter candidate 
will depend
%depends 
on the couplings. Motivated by the null observation of any baryon- or large lepton-number violating processes, we focus on models that are $ R $-parity conserving. Thus there is a  dark matter candidate in the dark sector if 
\begin{equation} 
m  < M_{\rm SUSY},
\end{equation} 
 where $ M_{\rm SUSY}$ is the mass scale of the superpartners in the visible sector.
 
For the hidden sector to be approximately supersymmetric, the breaking must be sufficiently small, and gravity mediation must be subdominant. This imposes the requirements (note that we use $ \varepsilon $ schematically here, though we give it an explicit meaning in the context of a model later)
\begin{equation} \label{eq:mrat}
m_{3/2} \ll m   \ll \varepsilon  M_{\rm SUSY} \,,
\end{equation}
where $m_{3/2}$ is the gravitino mass. 
Under these conditions, we call the dark sector {\em very supersymmetric}. This leads to the spectrum shown schematically in the right panel of Fig.~\ref{fig:sectors}.

The hidden sector will have a lightest $ R $-even and $ R  $-odd state, which we label $ \phi $ and $\widetilde{\psi}$, respectively. While $R$-parity protects $\widetilde{\psi}$ from decaying to SM particles, the mass hierarchy of Eq.~\eqref{eq:mrat} enables it to decay via couplings to the gravitino, and therefore $ \widetilde{\psi} $ is only an approximately stable dark matter candidate. These decays of  $ \widetilde{\psi}  $  proceed into the gravitino and either an on-shell or off-shell $ \phi $, depending on the mass hierarchy between the $\phi $ and $ \widetilde{\psi} $ states: $\widetilde{\psi} \rightarrow \widetilde{G} \of{ \phi^{(*)} \rightarrow f \bar{f} }$. The decay width of $ \widetilde{\psi}  $ is suppressed by either the small mass-splitting or by the small coupling to the SM. As a result, $ \widetilde{\psi} $  is long-lived on cosmological scales and is a viable DM candidate. 

The $ R $-even state, $ \phi $, will decay into SM particles via the $\varepsilon$ portal to the visible sector,
\begin{equation} 
  \Gamma _{ S \rightarrow {\rm SM}} \propto \varepsilon  ^2 m _{\rm} .
\end{equation} 
Effectively, the important interactions between the hidden and visible sector can arise from this decay. 
 
In essence, the setup we have just described contains all the ingredients of a co-decaying dark matter setup: Two nearly degenerate particles, interacting with each other, and one of them decaying into SM particles. The dynamics in the dark sector, and which process sets the relic abundance of the dark matter, both strongly depend on the size of these decays. 

Regardless of the existence of a dark matter candidate, this setup will generically alleviate the gravitino abundance problem. Since the lightest states in the dark sector are long-lived, the dark sector will dominate the energy density of the universe after they become non-relativistic. This leads to an early period of matter domination which has important effects on small scale structure~\cite{Dror:2017gjq}. Once the dark sector decays, it reheats the SM bath, diluting existing relics. The cosmological gravitino problem---that gravitinos are typically overproduced in the early universe~\cite{Moroi:1993mb}---is thus addressed.

%%%%%%%%%%%%%%%
\section{Co-decaying dark matter}
\label{sec:co-decay}

In this section, we briefly review the dynamics of decaying dark sectors with emphasis on the co-decaying regime. A detailed study of the thermal history is contained in Ref.~\cite{Dror:2016rxc}. Co-decay takes place when the lowest lying states in a hidden sector are degenerate and in thermal contact, and part of that sector is unstable.
The density of the stable states will track that of the unstable states as they decay away,
until the interactions between the two freeze out, and the stable states remain as relics. 
Co-decaying DM will be a generic feature of nearly supersymmetric dark sectors,  in which $R$-parity stabilizes one or more of the light states, while the others decay via a weak portal to the MSSM sector.  %

The relic density for a co-decaying sector can be given in terms of the standard sudden freeze-out approximation,
\begin{equation}
  \Omega  \simeq \bigg(\frac{10.75}{{g_{\star , f}}} \bigg)^{1/2}\bigg( \frac{x_f}{20} \bigg)\bigg( \frac{10^{-26} \rm cm^3/s}{\left\langle \sigma v \right\rangle _f} \bigg)\,, 
  \label{sf1}
\end{equation}
where $g_{\star}$ is the number of relativistic degrees of freedom,  $x\equiv m/T$ with $m$ the mass of the degenerate states and $T$ the SM bath temperature, and ${\left< \sigma v \right>_f}$ is the thermally averaged annihilation cross-section between the stable and unstable states. Here, subscripts $f$ refer to quantities evaluated at the time of freezeout, $T=T_f $.

For $s$-wave scattering of degenerate states, the cross-section approaches a constant, and therefore the thermally averaged cross-section is proportional to the average velocity. In terms of the on-threshold cross-section $\sigma$,  
\begin{equation}
\langle \sigma v \rangle_f =  \frac{4}{\sqrt{\pi}}\frac{\sigma}{ \sqrt{\smash[b]{x_f^{\prime}}}}\,,
\end{equation}
where $x^\prime=m/T^\prime$ and $T^\prime$ is the temperature of the dark sector.
Using this, the relic abundance of Eq.~\eqref{sf1} is 
\begin{equation}
  \Omega  \simeq 0.27 \times \bigg(\frac{10.75}{{g_{\star , f}}} \bigg)^{1/2}\bigg( \frac{x_f }{20} \bigg)\bigg( \frac{2\times 10^{-36} \rm cm^2}{\sigma}\bigg) \sqrt{\smash[b]{x_f^\prime}}\,.
  \label{omega}
\end{equation}

If the decay $\Gamma$ of the unstable state is 
faster than the Hubble rate at $T^\prime=m$, $ \Gamma \gtrsim H_m$, then the dark sector will decay and freeze out while in equilibrium with the SM, in which case the thermal history resembles that of a WIMP-like thermal relic, and $x_f = x_f^\prime \simeq 20$. The phenomenology of such a scenario was studied in detail in Ref.~\cite{Kopp:2016yji}.

Another possibility is that in the early universe, the dark sector decouples from the SM bath. The heavier states freeze out and decouple, leaving just the lightest states in the sector.  
Assuming the decays of the unstable light states occur out of equilibrium, $ \Gamma \ll H_m $, the comoving number density does not deplete when the nearly degenerate light states become non-relativistic. Instead, the exponential suppression of the number density is delayed until the unstable states decay. After decays begin (at a later time), the stable population continues to follow the unstable population until the system can no longer maintain chemical equilibrium. At this point, the stable population freezes out and the unstable particles decay away.  The relic density of the stable particles will depend on the cross-section, $\sigma$, between the stable and unstable states, as well as the lifetime of the decaying states, $\Gamma^{-1}$. 

The evolution of the dark matter density of the stable state $\widetilde \psi$ and the long-lived state $\phi$ are given by the Boltzmann equations:
\begin{equation}
 \def\arraystretch{1.5}
\begin{array}{l}
\dot{n}_{\widetilde \psi} + 3 H n_{\widetilde \psi} =- \langle \sigma v \rangle_{\widetilde \psi \widetilde \psi \to \phi \phi} (n_{\widetilde \psi} ^2 - n_{\phi}^2)  + [3\to2]\,,\\ 
\dot{n}_{{\widetilde \psi} +{\phi}} + 3 H n_{{\widetilde \psi} +{\phi}} =  - \left(  \langle \Gamma_{\phi} \rangle_{T^\prime} n_{\phi} -  \langle \Gamma_{\phi} \rangle_T n^{{\rm eq}}_T\right)   + [3\to2] \,, \\
\dot{\rho}_{{\widetilde \psi} +{\phi}}+ 3 H ({\rho}_{{\widetilde \psi} +{\phi}}+ P_{{\widetilde \psi} +{\phi}}) = -  m \Gamma_{\phi} \left(n_{\phi} - n^{{\rm eq}}_T \right)  \,, \label{BE}
\end{array}
\end{equation}
where $\langle \Gamma_\phi \rangle_{T(T^\prime)}$ is the thermally averaged decay rate at the given temperature. Here $n,~\rho$ and $P$ are the number density, energy density  and pressure, respectively. These equations can be solved, along with the Friedman equations, to obtain the relic abundance of $\widetilde{\psi}$, along with the entropy dump of the out-of-equilibrium $\phi$ decays, for given mass and interactions. We solve these equations numerically for the explicit model given below.

Useful analytical approximations that roughly capture the behavior of the relevant quantities can be found as follows. 
For an estimate of the relic abundance in terms of the parameters, one must determine the temperatures of the dark sector and SM at the time of freeze-out. The dark sector undergoes several stages of  evolution:
\begin{enumerate}
\item {\em Cannibalization:}	When the dark sector becomes non-relativistic, it undergoes cannibalization as the result of $3\to2$ processes.
\item {\em Non-relativistic cooling:} After $3\to2$ processes freeze out, the dark sector will redshift as a non-interacting non-relativistic particle. 
\item {\em Decay:} When the age of the universe reaches the average lifetime of the decaying states, i.e.~$\Gamma = H$, the dark matter particles will begin to decay away. 
\item {\em Freezeout:} The decays will stop depleting the DM abundance when the dark-sector interactions freeze out. 
\end{enumerate}
The temperatures at freeze-out can be obtained by evolving the dark sector density and temperatures through these different stages in the dark matter evolution. 

Using conservation of entropy and the second law of thermodynamics, one can show that the number density of the dark sector after it has become non-relativistic up until the onset of decays ($m< T^\prime <T_\Gamma^\prime$) is
\begin{equation} 
n^\prime =g^\prime\left(\frac{m T^\prime}{2\pi}\right)^{3/2} e^{(\mu^\prime -m)/T^\prime}=  \frac{T^\prime}{m- \mu^\prime+\frac{5}{2} T^\prime} \xi s, 
\label{nG}
\end{equation} 
where $\xi$ is the ratio of dark to SM degrees of freedom at decoupling, and $s$ is the entropy density of the photon bath~\cite{Dror:2016rxc}. During this period the dark matter can come to dominate the energy density of the universe, and therefore the Hubble scale has important contributions from both the energy density of the radiation and the non-relativistic dark matter:
\begin{equation} 
  H^2 = \frac{\rho_{\rm SM}+ m n^\prime}{3 m^2_{\rm pl}}\,. \label{H}
\end{equation} 

We refer to the temperatures at which cannibalization ends as $T_c$ and $T_c^\prime$. 
These temperatures can be found from the instantaneous freeze-out approximation,
\begin{equation}
n^2_{c}  \left< \sigma v^2\right>_{3\to2}   \simeq {H_c},
\label{xc}
\end{equation}
and from Eq.~(\ref{nG}), using $\mu_c=0$. Here subscripts $c$ refer to quantities evaluated at $T=T_c$. 

Next, the dark sector redshifts as a decoupled non-relativistic species until decays begin, roughly when 
\begin{equation} 
\Gamma = H_\Gamma.\label{xG}
\end{equation} 
The temperatures and chemical potential are simply give by
\begin{align} 
  x _\Gamma^\prime  &\simeq  \left( \frac{ g _{ \ast ,c } }{ g _{ \ast , \Gamma } } \right) ^{ 2/3}\frac{ x _c^\prime }{ x _c ^2 } x _\Gamma ^2 \,, \label{xGp}\\
  \frac{ \mu ' _\Gamma }{ m } &\simeq 1 - \frac{ x _c ' }{ x _\Gamma '}\,. \label{muGp}
\end{align} 
Eqs.~(\ref{xGp})--(\ref{muGp}) can be solved for $x_\Gamma$ and $x_\Gamma^\prime$.

Finally, the temperature at freeze-out can be determined using the instantaneous freeze-out approximation for the $2\to 2$ annihilations, and the number density for a decaying particle. Further details can be found in the original work~\cite{Dror:2016rxc}. We find that
\begin{equation}
x_f \simeq \frac{2  }{\sqrt{\Gamma_B / H_m} }\log^{1/2} \frac{\frac{2}{ \sqrt{ \pi }}\frac{s_m}{H_m} \xi  \sigma  }{ x _f \sqrt{ x_f^{\prime }} x_\Gamma^\prime (1-\frac{\mu^\prime_\Gamma}{m} + \frac{5}{2x_\Gamma^\prime} ) } \,, \label{xf}
\end{equation}
and the dark sector approximately continues to redshift until freeze-out, at
\begin{equation} 
x_f^\prime \simeq x_\Gamma^\prime \left(\frac{x_f}{x_\Gamma}\right)^2 \label{xfp}\,.
\end{equation} 

Eqs.~(\ref{omega})--(\ref{xfp}) can be solved for the relic abundance when the decays occur out-of-equilibrium.  As stated above, however, in presenting our results in the next sections, we use the full numerical computation.

%%%%%%%%%%
\section{Explicit model}
\label{sec:model}

We now present an explicit realization of a very supersymmetric hidden sector model and explore all the co-decaying parameter space. The model forms a useful benchmark to demonstrate in detail how co-decay occurs within supersymmetric theories, including both in- and out-of-equilibrium decays.

Consider the MSSM coupled via a small coupling $\varepsilon$, to a chiral superfield $\Phi$ with self-coupling, $\kappa$, and mass parameter, $m$. The
 superpotential is
\begin{align}
W = W_\mathrm{MSSM} + \frac{1}{2} m\Phi^2 + \frac{1}{3}  \kappa \Phi^3 + \varepsilon \Phi H_u 
 H_d, \label{superp}
\end{align}
where 
\begin{equation} 
W_\mathrm{MSSM}  = \mu H_u 
H_d + y_u H_u Q \bar{u} - y_d H_d Q \bar{d} - y_e H_d L \bar{e}\,.
\end{equation} 
We follow the notation conventions of Ref.~\cite{Martin:1997ns}.
Note that a similar Lagrangian is considered in the NMSSM~\cite{Nilles:1982dy,Frere:1983ag,Derendinger:1983bz} and in stealth-supersymmetry models~\cite{Fan:2011yu,Fan:2012jf,Fan:2015mxp}, but here the $\Phi$ field is not responsible for generating the $\mu$-term and the collider signatures are not stealthy.

Contained in the $\Phi$ superfield are the $R$-even complex scalar ${\phi = \frac{1}{\sqrt{2}}\left( \phi_R + i \phi_I \right)}$ and the $R$-odd fermion ${\widetilde{\psi}}$. The fermion, ${\widetilde{\psi}}$, will be taken to be the lightest $R$-odd particle (up to the gravitino).  After the scalar Higgses acquire their vacuum expectation values (VEVs), the portal interaction will induce mixing between the ${\widetilde{\psi}}$ and the MSSM neutralinos, and between the $\phi$ and the MSSM Higgses. Since we consider the portal to be very weak, $\varepsilon \ll 1$, the hidden sector mass eigenstates will contain a small amount of MSSM fields,  so we will continue to call the mass eigenstates by $\phi_R$, $\phi_I$, and ${\widetilde{\psi}}$. Full details of the mixing can be found in Appendix~\ref{sec:appA}.

All the ingredients for the co-decay mechanism are present in the superpotential Eq.~\eqref{superp}. The supersymmetric nature of the dark sector imposes approximate degeneracy between the fermion and scalar fields, $R$-parity stabilizes the fermion, the self-coupling  $\kappa$  induces interactions within the dark sector that keep it in equilibrium, and the small portal coupling, $\varepsilon$, will result in the scalars decaying to the SM. We review these ingredients here and leave the detailed calculations to the Appendix.

Mass splitting between the scalars and the fermion is generated by the mixing with the Higgs sector and via supergravity. A correction of ${\cal O}(\varepsilon)$ for the fermion mass is also generated, coming from the VEV of $\phi$ and the $\kappa \Phi^3$ interaction.  The masses of the hidden sector particles at this order are given by
\begin{align}
\begin{split}
m_{\phi_R}^2 = &  m^2 - \kappa \varepsilon v^2 \of{3 \frac{\mu}{m} - \sin\of{2\beta} }
+ \mathcal{O}( m m_{3/2})
, \\
m_{\phi_I}^2 = &  m^2 - \kappa \varepsilon v^2 \of{ \frac{\mu}{m } - \sin\of{2\beta} }
- \mathcal{O}( m m_{3/2})
, \\ 
m_{{\widetilde{\psi}}} = &  m - \frac{ \kappa \varepsilon v^2}{ m } \of{ \frac{\mu}{m } - \frac{1}{2}\sin\of{2\beta}}+\mathcal{O}( m_{3/2}).
\end{split}
\label{eq:masses}
\end{align} 
Note that in general the gravity-mediated effects are not precisely calculable without a complete model of the SUSY breaking sector; above we simply show the order of magnitude  expected in a generic model. 

For the  results in Section~\ref{sec:co-decay} to apply, the mass splitting between the fermion and scalars $\Delta m$ must be smaller than the temperature at freezeout,  $\abs{\Delta m} < T_f^\prime$ . If the mass splitting is larger, the model can still exhibit a viable dark matter candidate, but the parameters differ from those previously described.
 
For $m  < {m_h,m_H,m_A}$, the scalars decay to SM pairs via mixing with the MSSM Higgs bosons, 
\begin{align} \label{eq:phidecay}
\Gamma_{\phi_R  \to \rm SM} &\sim \varepsilon^2 \frac{v^2}{m_{h,H}^2} \Gamma_{h,H \to \rm SM}(m), \nonumber\\
\Gamma_{\phi_I  \to \rm SM} &\sim \varepsilon^2 \frac{v^2}{m_{A}^2} \Gamma_{A \to \rm SM}(m) ,
\end{align} 
where $\Gamma_{h,H,A \to \rm SM}(m ) $ is the decay width of the scalar Higgs evaluated with mass $m_{h,H,A} =  m $. For heavier dark scalar masses, the dark scalars can also decay directly in a pair of MSSM Higgs scalars. More precise decay widths are given in Appendix~\ref{sec:appB}. If the scalars are degenerate, we can define the complex scalar decay rate, $\Gamma_\phi = (\Gamma_{\phi_R} + \Gamma_{\phi_I})/2$.

 The trilinear superpotential coupling, $\kappa \Phi^3$, generates self-interactions within the dark sector. The ${\widetilde{\psi}} {\widetilde{\psi}} \to \phi \phi$ annihilation cross section is given by 
\begin{equation} \label{eq:XS}
 \sigma = \frac{25}{72\pi} \frac{ {\kappa}^4 }{ m^2 }\,,
\end{equation} 
and is responsible for maintaining chemical equilibrium within the dark sector until freezeout. 

The interaction also generates $3\to 2$ annihilations  that can cannibalize the dark sector when non-relativistic. There are many different possible $3\to2$ processes, involving different numbers of scalars and fermions. The temperature when cannibalization ends 
can be estimated from the total $3\to2$ thermally averaged rate, averaged over all fermions and scalars that can appear in the initial and final state. We find this to be
\begin{equation} 
\left< \sigma_{3 \to 2} v^2 \right> = \frac{25 \sqrt{5}}{1024 \pi} \frac{\kappa^6}{m^5}\,. 
\end{equation} 
The temperature of freeze-out of the $3\to2$ processes can then be determined by the instantaneous freeze-out condition, $n_{{\widetilde{\psi}},\phi}^2 \left< \sigma_{3 \to 2} v^2 \right>  = H$.

\begin{figure}[t!]
\includegraphics[width=.48\textwidth]{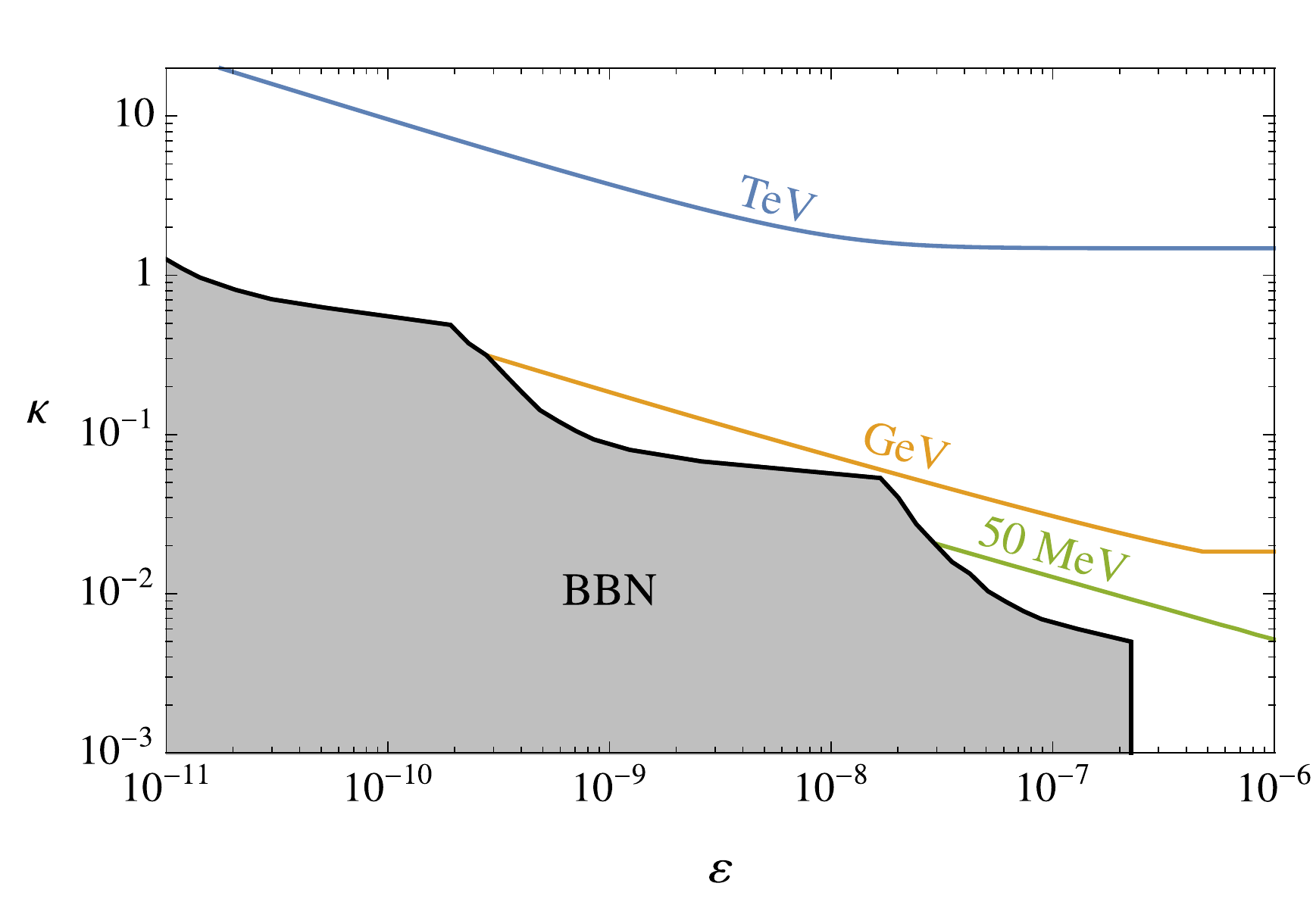}
\caption{\label{fig:phaseplot}
Setting the dark matter relic abundance via the co-decay mechanism, for various values of $m$ indicated on the solid curves. The shaded gray region indicates decays that occur past neutrino decoupling and is thus excluded by BBN measurements. 
}
\end{figure}

Assuming the mass splitting is small, the relic abundance depends on $m$, $\kappa$, and $\varepsilon$, with mild dependence on the standard MSSM parameters. In presenting our results throughout this paper, we choose a representative set of values for these, 
\begin{align}
\begin{split}
& \mu = 1 \ {\rm TeV}, \ m_A = 7000\ {\rm GeV},  \\
& \cos\of{\alpha-\beta} = 0, \ \tan\beta = 10, 
\end{split}
\label{eq:bmvals}
\end{align}  
but note that the qualitative features are unchanged for variations in these values.  In Fig.~\ref{fig:phaseplot} we show the $\kappa$ versus $\varepsilon$ relation that gives the observed dark matter relic abundance, for several values of $m$, as indicated by the solid colored curves.  The gray shaded region indicates where decays occur after neutrino decoupling and is thus excluded by Big Bang Nucleosynthesis (BBN); see discussion below.

%%%%%%%%%%%
\section{Constraints}\label{sec:constraints}
Having presented the model,  we move to discuss the constraints on the parameter space. The important parameters are $ m _{3/2} $, $ m $, $ \varepsilon$ and $ \kappa $, 
with the supersymmetric parameters taken illustratively as the benchmark values in Eq.~(\ref{eq:bmvals}). 
To obtain the correct relic abundance, we take $\widetilde{\psi}$ to make up the entirety of dark matter, which fixes the value of $\kappa$ in terms of $\varepsilon$ and $m$, as demonstrated in Fig.~\ref{fig:phaseplot}. The resulting parameter space is presented in Fig.~\ref{fig:constraints} in terms of the $(\varepsilon, m)$ parameter space. Below we describe the various constraints of each region.

%%%%%%%%%%
\subsection{Lifetime of $\phi$}\label{ssec:lifetimephi}
\begin{figure}[t!]
\includegraphics[width=\linewidth]{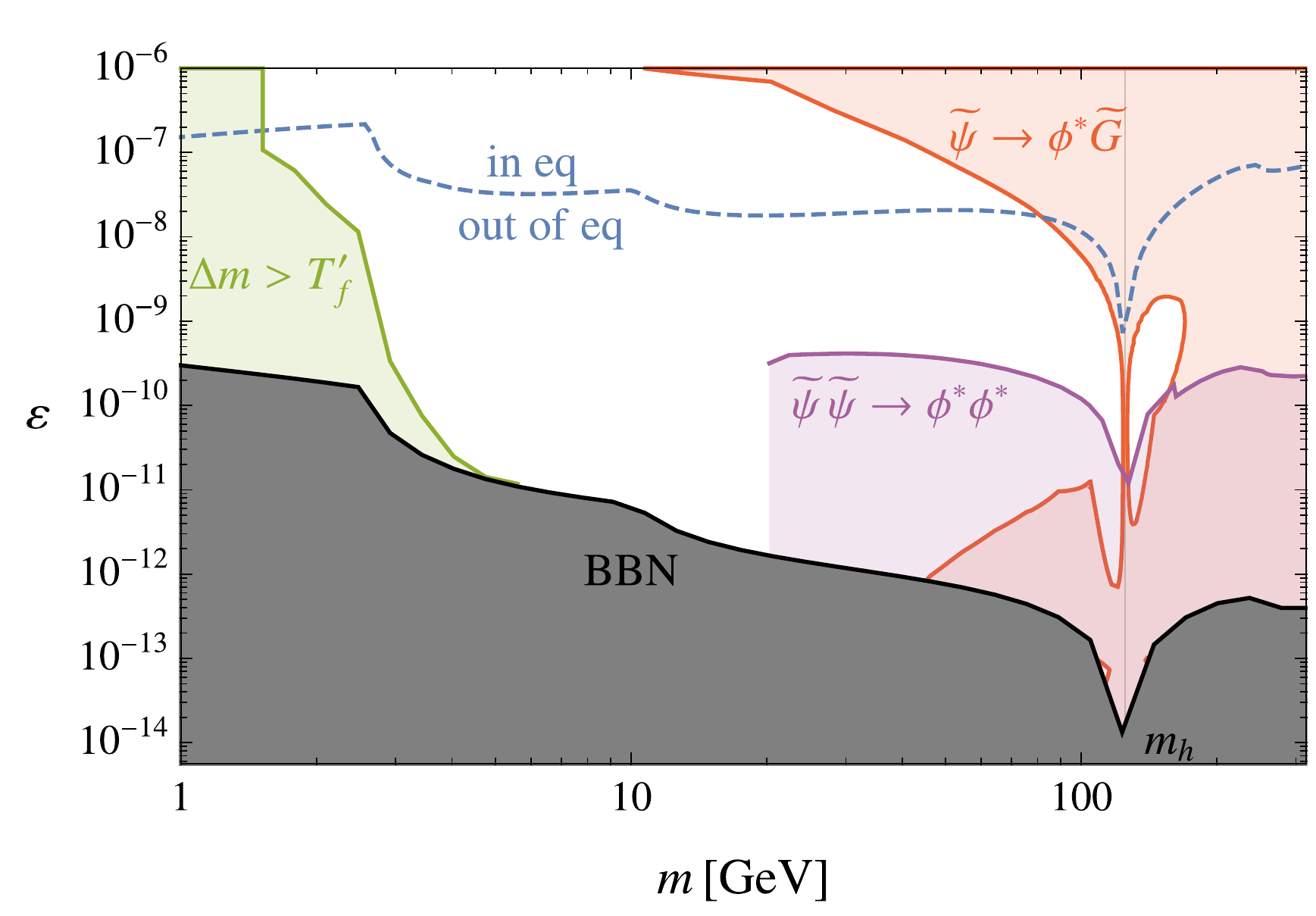}
\caption{Constraints on the parameter space. The gray shaded region indicates where decays occur after neutrino decoupling and is excluded by BBN. The blue dashed curve indicates the transition between in-equilibrium decays (above this curve) and out-of-equilibrium decays (below the curve). The green shaded region shows where the mass splitting (ignoring gravity contributions) is larger than the freeze-out temperature, in which case the relic abundance analysis is altered from that presented in this work. The red (purple) shaded region indicates where telescope experiments searches would have observed DM decay (annihilations).
}
\label{fig:constraints}
\end{figure}

The decays of the scalars can occur either in- or out-of-equilibrium. If either scalar has a sufficiently large decay rate [see Eq.~\eqref{eq:phidecay}] it will bring the entire sector into equilibrium with the SM. The  differentiation between these two regions can roughly be estimated by comparing the decay width to the Hubble parameter when the temperature is of order the particle's mass, 
$ \Gamma_\phi \simeq 0.1 H (  m ) $. 
This distinction is represented by the blue dashed curve in Fig.~\ref{fig:constraints}: above (below) this curves, $\phi$ decays roughly occur in (out) of equilibrium.

  If the decays are active during BBN, this will dump entropy into the SM as well as induce photodisintegration of light nuclei. Both these processes are heavily constrained by measurements of the ratios of abundances of light elements today~\cite{Lindley:1984bg}. 
To this end, we require that both scalars decay before neutrino decoupling, $ T\simeq {\rm MeV} $. The resulting constraint is indicated by the shaded gray region labeled `BBN' in Fig.~\ref{fig:constraints} (and by the shaded gray region in Fig.~\ref{fig:phaseplot}).

%%%%%%%%
\subsection{Lifetime of $\widetilde{\psi}$}
\label{sec:fermionDecay}
Requiring the mass splitting between the fermion and the scalars of the dark sector be small imposes $m_{3/2} \ll m$. Thus, the gravitino is the LSP, and $R$-parity can only render the dark matter candidate $\widetilde{\psi}$ approximately stable. The light gravitino induces the decays of the dark matter:
three-body decays $\widetilde{\psi} \rightarrow\widetilde{G} \of{\phi^* \rightarrow f \bar{f} } $,
and two-body decays $\widetilde{\psi} \rightarrow \widetilde{G}\phi $ (if $m_{\widetilde{ \psi} } > m_{3/2} +m_\phi$).  The decay rates are given by
\begin{align}\label{eq:psidecay}
\begin{split}
	\Gamma_{\widetilde{\psi },{\rm 3-body}} = & \frac{m^5}{1152\pi^2M_{\rm Pl}^2 m_{3/2}^2 } \frac{ \Gamma_\phi }{  m}\,,\\
	\Gamma_{\widetilde{\psi },{\rm 2-body}} = & \frac{ m^5 }{ 48 \pi M_{\rm Pl}^2  m_{3/2}^2 } \of{ \frac{ \Delta m^2 }{ m^2 }}^4\,.
	\end{split}
\end{align}
These decay rates are observable both cosmologically and today. Decays of particles between recombination and reionization will dump energy into the SM bath which can induce distortions to the cosmic microwave background (CMB)~\cite{Slatyer:2016qyl}. Furthermore, unstable dark matter decay in DM-rich regions can be observed using telescope experiments, placing a constraint on its decay rate~\cite{Liu:2016ngs,Essig:2013goa}. In general, both these probes depend on the final state products and the mass of the decaying particle. Which probe is most stringent depends on the mass of $ \widetilde{\psi}  $ and the particular decay mode, however we find that, across the range of interest, constraints from the isotropic gamma ray background measured by FERMI-LAT~\cite{Ackermann:2014usa} are the most powerful~\cite{Liu:2016ngs}.  %on whether the $  2 $ or $ 3 $-body decay above dominates. 

The dominant fermion decay rate depends on the mass splitting in the dark sector. 
From Eq.~\eqref{eq:masses}, the splitting between masses squared of the hidden sector fermion and scalars is 
\begin{align}
\Delta m^2 \equiv & m_{\widetilde {\psi}}^2 - m_{\phi_{R,I} }^2= \pm  \kappa \varepsilon v^2 \of{\frac{\mu}{m}} + \mathcal{O}\of{m m_{3/2}}\,,
\end{align}
with the last term above indicating the gravity mediated effects, which are not precisely calculable without a complete model of the SUSY breaking sector. However, there are calculable unavoidable contributions from anomaly mediation. Given the dimensionful coupling $m$ in the superpotential, including the conformal compensator ($\Phi_c = 1 + m_{3/2} \theta^2$) in the interaction, generates a $B$-type term,
\begin{equation} 
\int d^2\theta \left( m \Phi_c \Phi \Phi \right)\to m m_{3/2} \phi \phi\,. 
\end{equation} 
Additional terms of the same order will arise from the supergravity Lagrangian~\cite{Fan:2012jf}.   

Generically, the 2-body decay is fast if kinematically allowed; it is shut off if
\begin{align}
\Delta m^2 < & 2 m m_{3/2}.\label{eq:no2body}
\end{align}
The irreducible contributions are of order this size, but we require that the sum of these contributions add to satisfy Eq.~\eqref{eq:no2body} such that the 2-body decay is turned off and the 3-body decay dominates.

The dominant constraints in our mass region of interest are then from decays to $ \widetilde{ \psi } \rightarrow \widetilde{G} \bar{b} b $ and related fermionic channels (we sum all relevant channels when estimating the constraints). Such decays have similar kinematics to those of $ {\rm DM} \rightarrow \bar{b}b  $ decays, which have been well studied in the literature and we use the results of Ref.~\cite{Liu:2016ngs} to estimate the constraints. Noting that the 3-body decay rate (denoted here simply by $\Gamma_{\widetilde{\psi}}$) is inversely proportional to the square of the gravitino mass---namely that $\Gamma_{\widetilde{\psi}} m_{3/2}^2$ is constant---the lifetime constraint can be written as a lower bound on the gravitino mass $m_{3/2}$ in terms of the other parameters:
\begin{align}
m_{3/2} > \of{  \frac{  \Gamma_{\widetilde{\psi}} \,m_{3/2}^2}{  \Gamma _{ {\rm lim}} } }^{1/2}\,.
\label{eq:m32lower}
\end{align}
where $ \Gamma _{ {\rm lim}} $ is the observed limit, and is a function of the dark matter mass.

If $m_{3/2}$ is too large,  a large mass splitting between the hidden sector particles will be induced, which renders the results of Section~\ref{sec:co-decay} inapplicable. For simplicity, we therefore choose to focus on the region of parameter space where the splitting is smaller than the dark sector temperature at freeze-out, $\Delta m < T_{f}'$.  
To achieve this without too much tuning, we  require
\begin{align}
m_{3/2} < T_f'\,. \label{eq:m32upper}
\end{align}

Combined, to satisfy both Eqs.~\eqref{eq:m32lower} and \eqref{eq:m32upper}, we impose that
\begin{align}\label{eq:joint}
T_f' > \of{  \frac{  \Gamma  _{ \widetilde{ \psi } }\,m _{3/2}^2}{  \Gamma _{ {\rm lim}} } } ^{1/2}\,.
\end{align}
As long as this constraint is satisfied, there exists a value of $m_{3/2}$ where the mass splitting is sufficiently small, and the 3-body decay rate is sufficiently slow. The red shaded region in Fig.~\ref{fig:constraints} is ruled out by failure to meet this constraint. The non-smooth behavior of the constraint for small $ \varepsilon $ is from rapid changes in $ T _f ' $ as cannibalization stays on the entire time or not depending on $ m $. 

\subsection{$\widetilde \psi \widetilde \psi $ annihilations}
In addition to slow dark matter decays, telescope experiments also have the ability to detect DM annihilations which cascade into SM states through $ \widetilde \psi \widetilde \psi  \rightarrow \phi   _{ R/I} \phi _{ R/I}   $. The produced final SM state depends on the mass, with the dominate state being $ \bar{b}b  $ for the range of interest. We  recast previous limits on direct DM annihilations~\cite{Elor:2015bho,Liu:2016ngs}, while rescaling the rates and kinematics to match a $ 2 \rightarrow 2 $ process to a $ 2 \rightarrow 4 $ process. Due to the inevitable velocity suppression in the annihilation cross-section, we use the constraints from diffuse emission instead of dwarf galaxies or the cosmic microwave background, which are typically stronger but have a larger velocity suppression in rates.

The constraints are shown in the purple shaded region in Fig.~\ref{fig:constraints}. Unfortunately, the recast constraints for the final states of interest are only available for dark matter masses above 10 GeV (which corresponds to $20 ~{\rm GeV} $ for our $ 2 \rightarrow 4 $ processes). Nevertheless since the experiments detect emission well below this threshold, it should be possible to extend these constraints in future work.

%%%%%%%%%
\subsection{Gravitino abundance}\label{ssec:gravitino}
A necessary requirement of  very supersymmetric sectors is the lightness of the gravitino; otherwise there will be inevitable large splitting in the hidden sector. Light gravitinos are a form of hot dark matter and cannot make up a substantial fraction of dark matter without being ruled out by the observation of structure at small scales. Unfortunately, gravitinos are typically overproduced in what is known as the cosmological gravitino problem~\cite{Moroi:1993mb} (see, e.g., Ref.~\cite{Hall:2013uga} for a recent review). 

Post-inflationary gravitino production occurs through two main processes: freeze-in production from superpartners dropping out of the SM bath and decays of the (second) lightest $ R  $-odd states in the spectrum into the gravitino. %outside of the  stable particles in the spectrum. 
The gravitino problem for very supersymmetric sectors is alleviated in two different ways. First, by assumption that the DM is roughly cosmologically stable, there may not be significant gravitino production from decays. As discussed in Section~\ref{sec:fermionDecay}, this decay rate can be heavily suppressed if $ \phi $ is heavier than $ \widetilde{ \psi} $. Second, if freezeout occurs while the hidden sector is out of equilibrium with the SM [$ \Gamma \ll H ( m ) $], this can result in an early period of (dark) matter domination. 
Once the decays become efficient, they reheat the SM bath, diluting existing relics, $ Y _f \simeq Y _\Gamma S _\Gamma / S _{ {\rm RH}} $,
where $Y \equiv n /s $ is the comoving number density.
 
The ratio of entropy after reheating to the onset of the decay can be estimated using energy conservation as (see, e.g.~\cite{Kolb:1990vq}))
\begin{equation} 
\frac{S_{\rm RH}}{S _\Gamma } \sim  \frac{m }{(\Gamma m_{\rm pl})^{1/2} } \frac{n_\Gamma}{s _\Gamma }\,.
\end{equation} 
Using conservation of entropy from the time cannibalization stops in the hidden sector until decays begin,
\begin{equation} 
\frac{S_{\rm RH}}{S _\Gamma }  \sim    \frac{T_c^\prime }{(\Gamma m_{\rm pl})^{1/2} }\,,
\label{eq:sratio}
\end{equation} 
where $ T _c ' $ is the temperature in the dark sector when cannibalization stops (namely $ 3 \rightarrow 2 $ process freezeout). Taking the extreme case that $ T _c ' \sim  m $ and the decay happens at $ T \sim {\rm MeV} $, one can see the entropy dump can be as large as $ \sim m / {\rm MeV} $. For a generic co-decaying model, this can in principle be as large $10^7$~\cite{Dror:2016rxc}, significantly diluting any pre-existing gravitino abundance.

\begin{figure} 
\begin{center} 
\includegraphics[width=8cm]{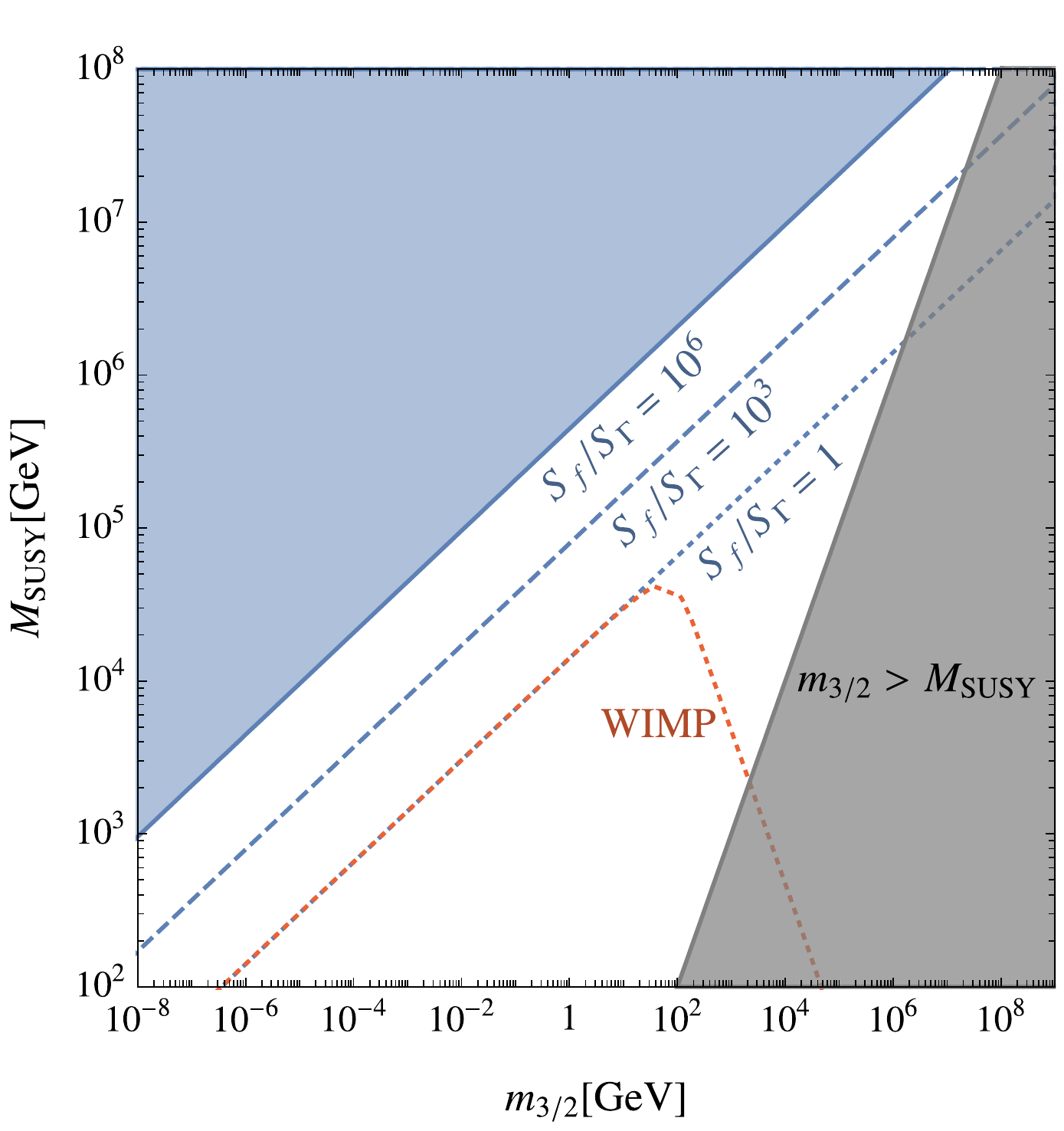} 
\end{center}
\caption{Constraints for different levels of dilution assuming $ T _{ RH} = M _{ {\rm SUSY}} $ (otherwise the constraints seep into the white region).}
\label{fig:gravitino}
\end{figure}

The initial gravitino abundance can be estimated using standard techniques~\cite{Hall:2013uga}. 
In Fig.~\ref{fig:gravitino} we show the constraints on the parameter space for different dilution factors assuming negligible production of gravitinos from the epoch of reheating after inflation. Note that co-decaying dark matter greatly reduces the gravitino abundance, allowing for relatively high scale inflation and superpartners well above the TeV scale. This opens up the possibility of gauge-mediated split supersymmetry~\cite{Cohen:2015lyp}.
Note that the entropy dump solution to the gravitino problem presented here does not depend on reproducing the relic abundance. In terms of the concrete model used here, the entropy dump has only a mild dependence on the parameters of the theory through the log-dependence of $T_c'$. 
 
%%%%%%%%%%%
\section{Discussion}\label{sec:discussion}

We have shown that very supersymmetric dark sectors, in which there is near degeneracy between bosons and fermions, can have distinct thermal histories which differ significantly from an ordinary WIMP. In much of the parameter space, the dark matter relic abundance is set by a co-decay process, where the decays of the scalar occur either in- or out-of-equilibrium. We presented a toy model which exemplifies all the features of such sectors, and studied the resulting constraints on the parameter space. Furthermore, we have demonstrated how very supersymmetric dark sectors can alleviate the gravitino problem via large entropy dumps, which are natural in such thermal histories. 

These supersymmetric dark sectors have distinct phenomenology and offer a new way to discover SUSY through the energy, lifetime, indirect detection, and cosmological frontiers.

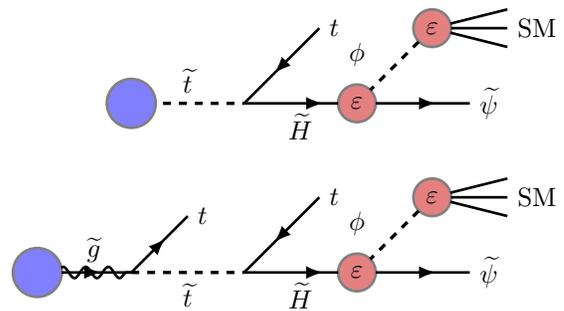
\begin{figure} 
\begin{center} \begin{tikzpicture} [line width=1]
    \draw[snar] (-0.5,1) -- (1,1) node[midway,yshift=0.3cm] {$ \widetilde{t} $};
    \draw[preaction={fill=white},fill=c1,opacity=0.5] (-.5,1) circle (0.325);
    \draw[decoration={markings,mark=at position 0.6 with {\arrow[very thick]{latex}}},postaction={decorate}] (2,2)node[right] {$ t $} -- (1,1) ;
    \draw[decoration={markings,mark=at position 0.7 with {\arrow[very thick]{latex}}},postaction={decorate}] (1,1) -- (2.5,1) node[midway,below] {$ \widetilde H $};
    \draw[decoration={markings,mark=at position 0.7 with {\arrow[very thick]{latex}}},postaction={decorate}] (2.5,1) -- (4,1) node[right] {$ \widetilde{\psi}  $};
    \draw[snar] (2.5,1)  -- (3.5,2) node[midway,xshift=-0.5cm,yshift=0.15cm] {$ \phi $};
    \draw[] (3.5,2) -- (4.5,1.75);
    \draw[] (3.5,2) -- (4.5,2) node[right] {SM};
    \draw[] (3.5,2) -- (4.5,2.25);
    \draw[preaction={fill=white},fill=c2,opacity=0.5] (2.5,1) circle (0.25cm);
    \draw[preaction={fill=white},fill=c2,opacity=0.5] (3.5,2) circle (0.25cm);
    \node[] at (2.5,1) {$\varepsilon $};
    \node[] at (3.5,2) {$\varepsilon $};

    \begin{scope}[shift={(0,-2.25)}]
      \draw[f] (-1.5,1) -- (-0.5,1) node[midway,yshift=0.3cm] {$ \widetilde{g}  $};
    \draw[v] (-1.5,1) -- (-0.5,1);

      \draw[snar] (-0.5,1) -- (1,1) node[midway,yshift=-0.3cm] {$ \widetilde{t}  $};
      \draw[f] (-0.5,1) -- (0.25,1.75) node[right] {$ t $};

%    \draw[snar] (0,0) -- (1,-1);
    \draw[preaction={fill=white},fill=c1,opacity=0.5] (-1.75,1) circle (0.325);
    \draw[decoration={markings,mark=at position 0.6 with {\arrow[very thick]{latex}}},postaction={decorate}] (2,2)node[right] {$ t $} -- (1,1) ;
    \draw[decoration={markings,mark=at position 0.7 with {\arrow[very thick]{latex}}},postaction={decorate}] (1,1) -- (2.5,1) node[midway,below] {$ \widetilde H $};
    \draw[decoration={markings,mark=at position 0.7 with {\arrow[very thick]{latex}}},postaction={decorate}] (2.5,1) -- (4,1) node[right] {$ \widetilde{\psi}  $};
    \draw[snar] (2.5,1)  -- (3.5,2) node[midway,xshift=-0.5cm,yshift=0.15cm] {$ \phi $};
    \draw[] (3.5,2) -- (4.5,1.75);
    \draw[] (3.5,2) -- (4.5,2) node[right] {SM};
    \draw[] (3.5,2) -- (4.5,2.25);
    \draw[preaction={fill=white},fill=c2,opacity=0.5] (2.5,1) circle (0.25cm);
    \draw[preaction={fill=white},fill=c2,opacity=0.5] (3.5,2) circle (0.25cm);
    \node[] at (2.5,1) {$\varepsilon $};
    \node[] at (3.5,2) {$\varepsilon $};
  \end{scope}
\end{tikzpicture}
\end{center}
\caption{Sample production modes for long-lived particles at colliders.}
\label{fig:production}
\end{figure}

At colliders, SUSY dark sectors will generically yield long-lived particles. If kinematically accessible, colored particles introduced to solve the hierarchy problem will be produced at the LHC. 
These will decay into the hidden sector through the small coupling $ \varepsilon $ between the dark and visible sectors, rendering these decays long-lived. A distinct feature of such sectors are decays both into the hidden sector and back again, which results in two displaced vertices, see Fig.~\ref{fig:production}. These vertices will have correlated decay lengths as they are both suppressed by the small coupling $ \varepsilon $. For the parameter space considered here, the lifetimes of $
\phi$ varies from a millimeter to effectively stable on collider scales. The lifetime for the LoSP is model dependent, but a similar range of lifetimes is expected. 
With the proposal of new experiments designed to discover long-lived particles such as MATHUSLA~\cite{Curtin:2018mvb}, FASER~\cite{Feng:2017uoz}, CODEX-b~\cite{Gligorov:2017nwh}, and AL3X~\cite{Gligorov:2018vkc}, the LHC could be primed to see one displaced vertex while the further detector could see the second, longer, displaced vertex.

Telescope experiments are also uniquely poised to observe very supersymmetric dark sectors via two distinct signatures. First, dark matter is predicted to inevitably decay to gravitinos 
and 
SM particles, with rates predicted by Eq.~\eqref{eq:psidecay}.  These decay rates depend on the couplings to the SM, the mass of the hidden sector, and the gravitino mass. In addition, one can also observe the annihilation process which governed the thermal freezeout as in Eq.~\eqref{eq:XS}. The annihilation rate depends only on the coupling within the hidden sector and its mass, which offers a complementary probe of the supersymmetric nature of the sector. An observation of both these signals 
could in principle be differentiated by the energy dependence of the final state.  

The above signatures persist for both the in- and out-of-equilibrium regions of parameter space. However, a unique prediction of the out of equilibrium co-decaying regime, is the possibility of a period of matter domination before the onset of BBN. This will produce enhanced power at small scales in the matter-power spectrum. As a result, compact dark objects can form, which could be seen through indirect detection~\cite{Erickcek:2015bda}, microlensing~\cite{2012PhRvD..86d3519L,Hezaveh:2014aoa}, astrometry~\cite{VanTilburg:2018ykj} and pulsar timing arrays~\cite{Baghram:2011is}. Unlike the collider signatures described above, enhanced power at small scales could be present even if SUSY is present well above the weak scale; this forms a unique opportunity to probe high scale supersymmetry.  

 {\bf Acknowledgements.} 
JD is supported in
part by the DOE under contract DE-AC02-05CH11231. The work of LHS is supported by the Zuckerman Foundation. The work of YH is supported by the Israel Science Foundation (grant No. 1112/17), by the Binational Science Foundation (grant No. 2016155), by the I-CORE Program of the Planning Budgeting Committee (grant No. 1937/12), by the German Israel Foundation (grant No. I-2487-303.7/2017), and  by the Azrieli Foundation. EK is supported by the Israel Science Foundation
(grant No. 1111/17), by the Binational Science Foundation
(grant No. 2016153) and by the I-CORE Program
of the Planning Budgeting Committee (grant No.
1937/12).
%%%%
%%%%

%\section{Appendix}
\begin{appendix}%{Technical details of the model}
\label{sec:appendix}

\section{Mixing of hidden sector with MSSM particles}\label{sec:appA}

Here we discuss the mixing of the hidden sector particles $\phi$ and $\widetilde{\psi}$ with the MSSM particles, in the model of Section~\ref{sec:model}.
The scalar potential is given by
\begin{align}
\begin{split}
	V =&  \abs{  \varepsilon\phi+\mu }^2   \of{  \abs{H_u}^2+\abs{H_d}^2 } \\ 
	&+  \abs{   \varepsilon  \of{  H_u \cdot H_d}  +\kappa^2+ m  \phi   }^2 \\%.
	& + m_{h_u}^2 \abs{H_u}^2  + m_{h_d}^2 \abs{H_d}^2 + \sqof{ b H_u \cdot H_d + \mathrm{c.c.} },
\end{split}
\end{align}
where $m_{h_u}^2$, $m_{h_d}^2$, and $b$ are soft squared-mass terms.
 
Once the scalars acquire VEVs, quartic terms in the potential generate mass corrections, and cubic terms induce mixing. 
Taking
$\phi = \frac{1}{\sqrt{2}} \of{ v_\phi + \phi_R + i \phi_I},$ 
one finds that
$\phi_R$ mixes with the CP-even MSSM Higgs fields $h$ and $H$, and $\phi_I$ mixes with the CP-odd Higgs $A^0$:
\begin{align}
\begin{split}
V \supset 
&
 \frac{\varepsilon v}{\sqrt{2}} m  \of{ 2 \frac{\mu}{m} + s_{2\beta}} \phi_R h  %\\
%&
 + \frac{\varepsilon v}{\sqrt{2}} m s_{2\beta}  \phi_R H \\
&
 + \frac{\varepsilon v}{\sqrt{2}} m \phi_I A^0,
\end{split}
\label{eq:Vphimix}
\end{align}
where $v = 246$ GeV, $\tan\beta\equiv v_u/v_d$, and we work in the alignment limit, where the light Higgs, $h$, is aligned with the direction of the VEV.
 
There are two values of $\phi_R$ that are local minima of the potential. Depending on the parameters, either may be the global minimum.
For $\varepsilon \ll  m^3 / v^2 \kappa \mu $, the possible VEVs are
\begin{align}
v_{\phi a} &= -\frac{  \varepsilon v^2 }{ \sqrt{2} m } \of{ \frac{\mu}{m} - \frac{1}{2} s_{2\beta} }; \label{eq:vphia}\\
v_{\phi b} &= -\sqrt{2} \frac{  m }{  \kappa }  - \frac{  \varepsilon v^2 }{ \sqrt{2} m } \of{  \frac{\mu}{m} + \frac{1}{2} s_{2\beta}  }.\label{eq:vphib}
\end{align}
For $\kappa \varepsilon < 0$, the global minimum is $v_{\phi a}$, which is $\varepsilon$ suppressed. For $\kappa \varepsilon >0$, the global minimum is $v_{\phi b}$, which is near $-\sqrt{2} m/\kappa$. To avoid complications associated with a large $\phi_R$ VEV, we choose ${\rm sign}\of{\kappa \varepsilon} = -1$, in which case the physical states are
\begin{align}
\begin{split}
\phi_{S_R} =  & \phi_R + \sqrt{2} \varepsilon v 
\sqof{ 
  \frac{m}{ 2 \of{m_h^2 - m^2 } } \of{
  							\frac{1 - t_\beta^2}{1+t_\beta^2}
							} h  \right. \\
 &	\left.	\hspace{1cm}	+ \frac{1}{ \of{ m_H^2 - m^2 } } \of{ \mu - \frac{t_\beta }{ 1 + t_\beta^2 } m } H 
  } \\
  & \hspace{1cm}  + \mathcal{O}\of{\varepsilon^2}   \end{split} \label{eq:SR}\\
\phi_{S_I}= & \phi_I + \varepsilon v  \frac{m }{\sqrt{2} \of{ m_A^2 - m^2 }  }  A^0 + \mathcal{O}\of{\varepsilon^2} \label{eq:SI}
\end{align}
with masses
\begin{align}
m_{S_R}^2 &= m^2 - \varepsilon \kappa v^2 \of{ \frac{  3 \mu}{ m}  - s_{2\beta}  }   + \mathcal{O}\of{ m_{3/2} m};\label{eq:mSR}\\
m_{S_I}^2 &= m^2 - \varepsilon \kappa v^2 \of{  \frac{\mu}{m}  - s_{2\beta}  }  + \mathcal{O}\of{ m_{3/2} m} \label{eq:mSI}\,.
\end{align}
 
The dark fermion $\widetilde{\psi}$ mixes with the MSSM neutralinos, as seen in the bilinear terms
%%%
\begin{align}
\begin{split}
\mathcal{L}_{\rm ferm.~mass} = & - \frac{1}{2}  \of{ m  + \sqrt{2} \kappa v_\phi } \psi \psi \\
& + \frac{ \varepsilon v }{ \sqrt{2} } \of{  \sqof{ s_\beta Z_{i3}^*  + c_\beta Z_{i4}^*  } \psi \widetilde{N}_i } \\
& -\frac{1}{2} m_{\widetilde{N}_i} \widetilde{N}_i \widetilde{N}_i  + \mathrm{c.c.}
%
%+ \of{  \mu + \frac{ \varepsilon v_{\phi} }{ \sqrt{2} } }Z_{i3}^* Z_{j4}^*  \widetilde{N}_i \widetilde{N}_j
%
\end{split}
\end{align}
with $v_\phi$ given by Eq.~(\ref{eq:vphia}) (which holds when $\varepsilon \kappa  <0$, as noted above). Writing the mass terms in matrix form,
\begin{align}
\mathcal{L}_{\psi,\widetilde{N}_i} = & -\frac{1}{2} \of{ \begin{array}{cc}  \psi  & \widetilde{N}_i  \end{array}} \mathbf{M}_{\psi}  \of{
\begin{array}{c}
\psi \\
\widetilde{N}_j
\end{array}
}
\end{align}
\begin{align}
\mathbf{M}_\psi = &
\of{
\begin{array}{cc}
  \of{ m + \sqrt{2} \kappa v_\phi } & - \frac{ \varepsilon v }{ \sqrt{2} }  \of{  s_\beta Z_{3j}  + c_\beta Z_{4j}  } \\
-\frac{\varepsilon v }{ \sqrt{2}}  \of{ s_\beta Z_{3i} + c_\beta Z_{4i}  }   &  m_{\widetilde{N}_i} \delta_{ij}
% \delta_{ij} \mu + \frac{ \varepsilon v }{ \sqrt{2}}  Z_{i3}^* Z_{j4}^* 
\end{array}
}
\end{align}
where $Z_{ij}$ is MSSM neutralino mixing matrix,
\begin{align}
\widetilde{N}_i = Z_{ij} \of{ \psi^0 }_j; \ \ \ \  \psi^0_i = \of{  \widetilde{B}, \widetilde{W}^0, \widetilde{H}_u^0 , \widetilde{H}_d^0 } .
\end{align}
Diagonalizing perturbatively, we find the dark fermion mass eigenstate is:
\begin{align}
\begin{split}
\psi - \frac{ \varepsilon v}{\sqrt{2}}  \sum_{i=1}^4 \frac{s_\beta Z_{3i} + c_\beta Z_{4i} }{m - m_{\widetilde{N}_i} } \widetilde{N}_i  + \mathcal{O}\of{\varepsilon^2},
\end{split}
\end{align}
and the dark fermion mass is
\begin{align}
\begin{split}
m_{\widetilde{\psi}} =& m - \varepsilon v^2 \ \frac{\kappa}{{m} } \of{\frac{\mu}{m} - \frac{1}{2} s_{2\beta}}  + \mathcal{O}\of{\varepsilon^2}.
\end{split}
\end{align}
From this and Eqs.~(\ref{eq:mSR}) and (\ref{eq:mSI}),  
the mass splittings in the hidden sector are
\begin{align}
\Delta m_{R,I}^2  \equiv m_{\widetilde{\psi}}^2 - m_{\phi_{R,I}}^2  = \pm \kappa \varepsilon v^2 \frac{\mu}{m}.%  - m m_{3/2}.
\end{align}
 
We emphasize that Eqs.~(\ref{eq:SR})--(\ref{eq:mSI}) rely on a perturbative expansion in the parameter $\varepsilon v^2 \kappa \mu / m^3$, which is small in only some of the parameter space we consider. To produce the plot in Fig.~\ref{fig:constraints}, rather than relying on this perturbative expansion to get analytic expressions, we numerically diagonalized the full mass matrix. We use the perturbative result only as a conceptual tool.

\section{Dark sector scalar decay rate}\label{sec:appB}

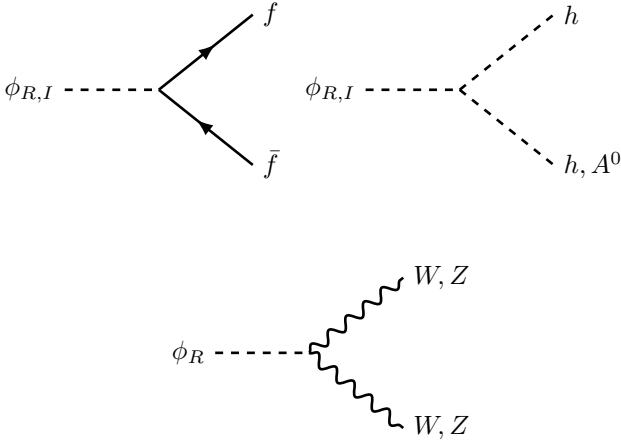
\begin{figure}
\begin{center} \begin{tikzpicture}[line width=1] 
  \draw[dashed] (0,0)  node[left] {$ \phi _{ R , I} $} -- (1.25,0);
  \draw[f] (1.25,0) -- (2.5,1) node[right] {$ f $};
  \draw[f] (2.5,-1) node[right] {$ \bar{f} $} -- (1.25,0) ;
  \begin{scope}[shift={(4,0)}]
  \draw[dashed] (0,0)  node[left] {$ \phi _{ R , I} $} -- (1.25,0);
  \draw[dashed] (1.25,0) -- (2.5,1) node[right] {$ h $};
  \draw[dashed] (1.25,0) -- (2.5,-1) node[right] {$ h, A ^0  $};
\end{scope}

  \begin{scope}[shift={(2,-3.5)}]
  \draw[dashed] (0,0)  node[left] {$ \phi _{ R } $} -- (1.25,0);
  \draw[v] (1.25,0) -- (2.5,1) node[right] {$ W,Z $};
  \draw[v] (1.25,0) -- (2.5,-1) node[right] {$ W,Z  $};
\end{scope}
\end{tikzpicture}\end{center} 
\caption{Decay modes of $\phi_R$ and $\phi_I$.}
 %(or $S_R$ and $S_I$). }
\label{fig:phidecay}
\end{figure}

The scalars $\phi_R$, $\phi_I$ decay to MSSM particles via the $\varepsilon$ portal.  These decay rates can be perturbatively calculated from the diagrams shown in Fig.~\ref{fig:phidecay}, using the mass-insertions in Eq.~\eqref{eq:Vphimix}. 
The total width, for a given mass, is the sum of partial widths to all SM final states for which $\sum m_i \leq m$, where the partial decays of the real scalar are,\begin{align}
\Gamma^{\phi_R}_{f\bar{f}} = & \frac{ \varepsilon^2 v^2 }{ 2 m^2 }  \sqof{  
\frac{\of{   2 \frac{\mu}{m} - s_{2\beta} }  }{  m^2 - m_h^2  }
+
\frac{ \of{ y^f_H  / y^f_h } c_{2\beta}   }{m^2 - m_H^2 }
}^2 \Gamma^h_{f\bar{f}}  \\
\Gamma^{\phi_R}_{hh} = & \frac{ \varepsilon^2 m }{  64 \pi }   \of{   2 \frac{\mu}{m} - s_{2\beta} }^2 \sqrt{ 1 - \frac{m_h^2 }{ m^2 } }  \\
\Gamma^{\phi_R}_{gg} = &  \frac{ \varepsilon^2 v^2 m_h \of{ 2 \frac{\mu}{m} - s_{2\beta} }^2 }{ 2 m^3 \of{ 1 - \frac{m_h^2 }{m^2}}^2 }  \Gamma^h_{gg} \\
\Gamma^{\phi_R}_{ZZ, WW} = &  \frac{  \varepsilon^2 v^2 \of{ 2 \frac{\mu}{m} - s_{2\beta}  }^2 }{ 2 m^2 \of{ 1 - \frac{m_h^2}{m^2} }^2 } \Gamma^h_{ZZ,WW} 
%
%
%\Gamma^{\phi_I}_{f\bar{f}} = &  \frac{\varepsilon^2 v^2 }{ 2 m^2 } \of{ 1 - \frac{ m_A^2 }{ m^2 } }^{-2}  \Gamma^A_{f\bar{f}}
\end{align}

For the imaginary scalar, 
\begin{align} 
\Gamma^{\phi_I}_{f\bar{f}}  = &  \frac{\varepsilon^2 v^2}{m_{\phi_I}m_A}\left(1-\frac{m_A^2}{m_{\phi_I}^2} \right)^{-2} \Gamma^A_{f \bar f}\,, \\
\Gamma^{\phi_I}_{hA}  & \simeq  \frac{\varepsilon^2 m_{\phi_I}}{64\pi}\frac{m_{\phi_I}^2-m_A^2}{2m_{\phi_I}^2} \,,
\end{align}
%
%with $\beta^X_Y \equiv \left(1-4m_Y^2/m_X^2\right)^{1/2}$ and 
where $\Gamma^{h/A}$ denotes the $ h $ or $ A $ decay rate (see, e.g., Ref.~\cite{Aoki:2009ha}) with $m_{h/A} \rightarrow m$.

\end{appendix}

\bibliography{SUSYDM}

\end{document}